# Governing rapid technological change: Policy Delphi on the future of European AI governance

Atte Ojanen [a][b], Johannes Anttila [a], Thilo H. K. Thelitz [a], Anna Björk [a]

[a] Demos Helsinki  [b] University of Turku

*The rapid advancements in artificial intelligence (AI) present unique challenges for policymakers that seek to govern the technology. In this context, the Delphi method has become an established way to identify consensus and disagreement on emerging technological issues among experts in the field of futures studies and foresight. The aim of this article is twofold: first, it examines key tensions experts see in the development of AI governance in Europe, and second, it reflects on the Delphi method's capacity to inform anticipatory governance of emerging technologies like AI based on these insights. The analysis is based on the results of a two-round Policy Delphi study on the future of AI governance with European policymakers, researchers and NGOs, conducted in mid-2024. The Policy Delphi proved useful in revealing diverse perspectives on European AI governance, drawing out a consensus that future-proof AI regulation will likely depend more on practical implementation and enforcement of legislation than on its technical specifics or scope. Furthermore, the study identified a desirability-probability gap in AI governance: desirable policy directions, like greater citizen participation, were perceived as less probable and feasible. This highlights a tension between desirable regulatory oversight and the practical difficulty for regulation to keep up with technological change.*

1. Introduction

The rapid development of artificial intelligence (AI) systems generate risks that require urgent policy responses, such as attributing liability for AI harms and managing concentration of market power. This technological development presents significant challenges for regulatory and legal frameworks to remain adaptable and resilient, as already noted by Collingridge (1982). The tension between technological progress and reactive policymaking is also reflected in the growing traction around the concept of anticipatory governance and its instruments (e.g., Guston, 2014; Fuerth, 2009, OECD, 2024). A key feature of anticipatory governance is the institutionalization of foresight mechanisms and capacities within policymaking to manage the risks of emerging technologies. Reflective of this, the European Union (EU) has built a significant digital policy apparatus over the past ten years through regulations such as the Artificial Intelligence Act (2024), Digital Services Act (2022), and the General Data Protection Regulation (2016). The EU has specifically positioned its regulation to be anticipatory and future-proof – to remain relevant and not become outdated despite technological progress (Ojanen, 2025a). Yet, this future-proofness is arguably already wavering as the European Commission (2025) has proposed deregulatory measures like the so-called Digital Omnibus in response to intensifying geopolitical competition.

In recent years the Delphi method – an anonymous survey designed to achieve expert consensus over multiple iterative rounds (Linstone & Turoff, 2002) – has been increasingly utilized to forecast technological development of AI in sectors like healthcare and education (e.g., Berbis et al., 2023; Güneş & Kaban, 2025). Yet, very few if any Delphi studies have focused on AI governance and its specific mechanisms, such as risk mitigation or transparency measures (see Alon et al. 2025). As such, the method's contribution to anticipatory governance in the context of AI remains underexplored. Instead of a traditional consensus-based Delphi study, we hypothesize that the Policy Delphi variation, based on exploration of potential conflicts between diverse policy perspectives, is better suited for the rapid progress and uncertainty characterizing AI development. Based on the results of a two-round Policy Delphi study on the future of AI governance with 29 European AI experts conducted in mid-2024, this article addresses the following research questions: 1) Substantively, what are the key tensions experts see in EU's AI governance, particularly in terms of its future-proofness?; and based on this 2) Methodologically, how can the Policy Delphi method inform the anticipatory governance of AI? This is done by analyzing experts' responses on probability, desirability and importance of different forward-looking AI policies, especially in the context of implementation of the EU AI Act. As such, the article contributes to understanding of how a Policy Delphi can challenge and validate assumptions about anticipatory governance and future-proof regulation in the uncertain and volatile context of AI.

In the second section of this article, we give a brief background on anticipatory governance of technology and AI, and how the Delphi method relates to it. The third section outlines the design of our Policy Delphi study on the future of AI governance, including the participant recruitment, round-structure and analysis. We present the results of the study in the fourth section, with focus on the different AI policy options that were thought to be important and desirable. Namely, we identify a desirability-probability gap in AI governance: the most desirable policies, such as greater citizen participation, were perceived to be less probable and feasible. These results are further discussed in the fifth section, focusing on the implications for anticipatory and future-proof AI regulation. Despite the diverse expert perspectives, a consensus emerged that future-proof AI regulation in Europe will depend more on practical implementation and enforcement of legislation than on its original scope or technical details. This challenges the prevailing EU consensus on the importance of technology neutrality and regulatory sandboxes for future-proofness, but rather emphasizes the credible commitment and resources of authorities for the regulation over time (see Ibáñez Colomo, 2022; Arnal, 2025). We conclude by reflecting on the potential of the Policy Delphi to inform the rapidly evolving landscape of AI governance.

## 2. Context and theory

The global landscape of AI governance[1] has been increasingly framed as a geopolitical race between major powers such as the US, China, and the EU (e.g., Maas, 2025; Smuha, 2021). This competition for technological leadership in terms of data, hardware, energy and talent is evident in new industrial policies and standard-setting for AI (Zúñiga et al., 2024). Yet, the AI race narrative is also tied to questionable future visions and expectations of artificial general intelligence (AGI), particularly the notion that achieving it first grants decisive strategic advantage over competitors (Ó hÉigeartaigh, 2025). This rhetoric, often espoused by leading AI companies, is also frequently used to limit regulatory oversight. Nevertheless, the speed of AI development coupled with its societal implications place traditional governance mechanisms under pressure. Perhaps nowhere else is the classic Collingridge dilemma more apparent: sociotechnical risks of AI systems are difficult to forecast until deployed at scale, at which point regulation becomes challenging (Collingridge, 1982). Similarly, the 'pacing problem' demonstrates that incremental laws and regulations which take multiple years to enact struggle to keep up with technological innovation (Downes, 2009).

---

[1] We define AI governance as the overall framework of rules and standards for AI, which also includes practices of private AI companies, whereas AI policy as its subset refers specifically to the policies and regulations created by public authorities (see Maas, 2025).

To resolve this conflict between information and time, the concept of anticipatory governance has gained prominence. Initially developed within the field of responsible innovation, anticipatory governance refers to the institutional capacity to manage emerging technologies and their societal impacts while effective intervention is still possible (Guston, 2014, 219). This involves integrating strategic foresight – the systematic analysis of possible, plausible, and preferable futures – into the policy design to monitor weak signals and assess policy performance in order to respond to changing circumstances proactively (Fuerth, 2009). As such, anticipatory governance seeks to systematically use methods like horizon scanning, trends analysis and scenarios to reduce technological risks and uncertainty in decision-making. As identified by Muiderman et al. (2020), approaches to anticipatory governance can range from probability-based risk reduction to more imaginative and participatory explorations of possible futures. The goal is not to predict the future with certainty, but to build institutional reflexivity in order to steer technological development towards societally desirable outcomes, instead of merely reacting to technologies. This entails proactive engagement with diverse stakeholders across policymakers, researchers, industry and citizens to leverage collective intelligence and align emerging technologies with societal needs (Panizzon & Janissek-Muniz, 2025).

In European legislation, these anticipatory principles have been operationalized through attempts to create future-proof regulation, such as the AI Act (see Ojanen, 2025a). Future-proofing is designed to make legislation adaptive against technological changes by introducing greater flexibility to it, so that regulation does not become obsolete as the technology evolves. The European Commission's Better Regulation Toolbox (2023, 176) includes flexibility and future-proofing as key legislative design considerations, highlighting less prescriptive and detailed measures to avoid locking in specific technical solutions. This is so that regulation can accommodate technological progress and be regularly updated as circumstances change without stifling innovation. A primary tool for this is technology neutrality, which stands for regulating the use and effects of technologies broadly, without discriminating against specific technologies. This neutrality should allow the legislation to adapt to technological changes over time, rather than being tied to certain outdated technologies (Koops, 2006; Greenberg, 2015). Aside from neutral legal language, future-proofing also requires the above-mentioned capacities for anticipating technological development and its effects through strategic foresight. In addition, it can also benefit from experimental methods such as ex-ante impact assessments, regulatory sandboxes and sunset clauses (Ranchordás & van 't Schip, 2020).

Despite the theoretical alignment, the application of foresight in technology policy remains piecemeal (OECD, 2024). One method for implementing participatory foresight required by anticipatory governance is the Delphi method. The Delphi method seeks to systematically distill the judgments of an expert panel into an impartial group consensus or forecast. This is achieved through a structured process of multiple anonymous questionnaires, with controlled feedback provided between rounds, allowing the experts to refine their answers (Linstone & Turoff, 2002). Yet, for policy issues which lack a single correct answer, a variation known as the Policy Delphi has been developed. Policy Delphi seeks to explore a complex policy issue by eliciting and clarifying the full range of expert opinions (Turoff, 1970; Ziglio, 1996). In contrast to the traditional Delphi's focus on forecasting and consensus, the Policy Delphi is designed to map disagreements and assumptions underlying uncertain policy issues and their governance options (Turoff, 2002). By providing a wider array of arguments and desirable policy options, the Policy Delphi is especially well-suited for identifying mutual priorities in the context of complex and uncertain policy problems, such as the governance of AI (Haynes et al., 2016).

Yet, there has been a lack of Policy Delphi studies focused specifically on AI governance and its mechanisms (see Alon et al. 2025). While numerous Delphi studies on AI have been published, these have been predominantly focused on sectoral use cases, like healthcare, education or manufacturing (e.g., Berbis et al., 2023; Güneş & Kaban, 2025; Demlehner et al., 2021). At most, existing Delphi studies have covered policy and ethical impacts of AI on the level of abstract principles. For instance, Stahl et al. (2023) use Delphi to map the main ethical issues raised by AI and their associated mitigation measures on a general level without connecting these to concrete policies in specific jurisdictions. Moreover, previous research has not evaluated the contribution of the Delphi method for anticipatory governance in the context of AI, characterized by rapid development, epistemic uncertainty and lack of scientific consensus (see Ojanen, 2025b). Therefore, this article asks how can the Policy Delphi method inform anticipatory governance of emerging technologies like AI, particularly in terms of future-proof regulation? The goal is two-fold: to yield novel insights on the substantive challenges of anticipatory AI regulation in the EU, and to assess the strengths and limitations of the Delphi method itself as a tool for this. We hypothesize that the Policy Delphi's ability to systematically map out the diverse expert perspectives on complex and uncertain policy issues makes it especially valuable for the rapidly evolving context of AI governance.

### 3. Method

This section covers how we designed, conducted and analyzed the results of a two-round Policy Delphi on the future of AI governance in mid-2024, consisting of 29 European AI experts. Across the two rounds, participants were provided with forward-looking questions on AI governance to assess their feasibility, desirability, importance and level of agreement as well as to justify their answers qualitatively. The first round sought to expose the issues experts disagreed most about in terms of AI governance, such as the risks of AI, gaps in the EU's AI governance, issues of future-proof regulation, the role of citizen participation and global cooperation in AI governance, and the need for industrial AI policy in Europe. Prior to this, we tested the questions we had identified from background research on AI governance in an expert roundtable in Brussels in October 2023. The workshop and discussion provided positive feedback and validated the general scope and direction of our research, forming a basis for the first Delphi questionnaire.

In April 2024, we sent out email invitations to fill out the questionnaire to 203 experts from academia, think tanks, industry, and EU institutions, namely the European Commission (EC) and Parliament (EP). In recruiting a panel from different groups, we followed Turoff's (2002) recommendation for Policy Delphis to be heterogeneous to capture different perspectives and improve representativeness. De Loë et al. (2016) identify this as a common practice for Policy Delphis, as well. The invited experts had demonstrated digital and AI policy expertise, for instance by publishing related journal articles or reports, or working in relevant units within DG CONNECT of the EC. The number of participants was a typical one, according to Turoff (2002, 82) and De Loë et al. (2016, 82). As table 1 demonstrates, we received most responses from academia (8 & 6), think tanks (7 & 7) and policymakers (5 & 5), but fewer responses from private business (2 & 1). This amounted to 22 responses in the first round, 19 responses in the second round, and 29 unique respondents in total. Overall, the survey design allowed us to investigate the experts' diverse views on key issues of AI governance in detail, although the low number of industry respondents should be taken into account when assessing certain questions.

*Table 1. Respondents to the Policy Delphi survey by expert category and round.*

| Expert category | Amount of respondents in round 1 | Amount of respondents in round 2 |
|---|---|---|
| Policymakers (EC and EP) | 5 | 5 |
| Academics (AI governance, law and ethics) | 8 | 6 |
| NGOs and think tanks | 7 | 7 |
| Industry | 2 | 1 |
| **Total** | **22** | **19 (12 i.e. 63% returning)** |

The questionnaire was implemented digitally on Google Forms. The first survey consisted of 18 items of different types: Eight items asked the respondents to assess the probability and desirability of certain statements on a five-point Likert scale, e.g., *"New participatory mechanisms for AI governance, such as citizen panels, will be established in the EU."* The Likert items ranged from very unlikely (1) to very likely (5), and from very undesirable (1) to very desirable (5). The other main category was ranking questions, where respondents had to rank pre-defined issues in terms of their importance or relevance in AI governance. While participants generally responded to every survey item, there was slight variance in response rates to some questions, with the final N= ranging between 21-22. Each question also included a text field for respondents to explain their reasoning and add comments, along with five general open text field questions. These qualitative items served not only to gather nuances of particular answers, but also to gauge how the questions were received, which helped shape the questionnaire for the subsequent round.

After the first Delphi round concluded, we analyzed the responses qualitatively and quantitatively. In multiple passes, we examined qualitative survey responses and how they related to the numerical answers to the Likert items to understand the expert reasoning holistically. We used thematic analysis to identify patterns, themes as well as opposing perspectives that emerged from the answers. This analysis also helped us notice when certain question types or phrasings may have hurt the validity of the numerical responses. The survey resulted in such a large volume of qualitative text responses that it is beyond the scope of this article to faithfully reproduce. Instead, we highlight the key themes from the analysis, illustrate them with select quotations, and contextualize the quantitative results accordingly. In terms of quantitative analysis, we began by applying the "agreement rating scale" used in Meskell et al. (2014) and inspired by O'Loughlin and Kelly (2004) to our data using spreadsheet formulas. As shown in table 2, we then calculated combined consensus levels for each Likert item using the same thresholds as in Meskell et al. (2014). Following this, we collected the scores, consensus levels and text responses into a "dataset" for each survey item, similarly to the methodology of De Loë (1995). For ranking questions, we simply took the sum of each item to generate an average ranking order, while discounting partial responses.

*Table 2. Percentage thresholds for consensus levels. Any value above the listed percentage threshold belongs to the associated level of consensus.*

| Category | High consensus level | Medium consensus level | Low consensus level |
|---|---|---|---|
| Combined (e.g., a 1 or a 2 on a Likert scale) | 80% | 70% | 60% |

The second Delphi round was designed based on the results of the first, focusing more on concrete policy options in AI governance. Therefore, many survey items asked the experts how important certain aspects of governance were, and whether certain avenues should or should not be pursued. In a break with established practice (cf. De Loë et al., 2016), we did not present the first round's results to the panel in full detail to refine, but rather referred to more general findings in the second-round survey. We chose this approach because of the large volume of first round data, the second round's purpose of elaborating on specific AI policy interventions and because the objective of the Policy Delphi was to expose a wide range of possible policy options instead of conforming to a group consensus (Turoff, 1970). The second round was consequently shorter than the first, with only 12 items. This time, there were nine survey items consisting of 5-point Likert scales and open text fields for comments, and two open text field questions. Because first round respondents pointed out that some of the survey's ranking items were difficult to answer and that they considered many options at an equal rank, we opted not to use further ranking items (see Appendix A for full tabulated results of these items). However, we still utilised Likert scale rating questions to measure perceived importance of AI policy options on an ordered scale.

Nineteen responses (excluding one partial response) were received for the second round, including twelve from experts who had also participated in the first round (see Table 1).[2] Contrary to conventional Delphi's, we allowed new experts to join the panel in the second round rather than restricting participation to only those who had responded initially, in order to maximize the diversity of perspectives. This design choice aligns with the core objective of a Policy Delphi, which prioritizes the exploration of diverse viewpoints, dissensus and potential policy pathways over achieving consensus as in traditional Delphi studies (see Linstone & Turoff, 2002). We therefore prioritized panel openness to ensure more inclusive evaluation of different AI policy options and priorities. This is also partly why second round respondents were not asked to refine first round responses, as new participants would not have been able to engage meaningfully. We did not observe any systematic differences in Likert item responses between returning and new respondents. After closing the second-round survey, we applied the same analysis tools as in the first round — thematic analysis and agreement rating scales. We then integrated the findings from both rounds, treating them as complementary sources of insight rather than as separate, sequential phases. Accordingly, below we present the results in an integrated manner across three thematic clusters.

## 4. Results

Overall, there was a strong consensus among the experts (R1Q1)[3] that the development of AI systems would pose significant transformations and risks to society (both ≥50%) within the next two decades, although more experts thought that the impacts would be transformative (36%) than that the risks would be extreme (18%). Multiple respondents noted that they were more concerned about near to mid-term societal risks of AI systems, rather than the long-term existential risk narrative. In terms of risks to democratic societies (R1Q2), panelists were especially concerned about concentration of power (55%), disinformation (46%), algorithmic discrimination (41%) as well as economic disruptions (32%). This was also reflected in how the experts saw the material AI infrastructure (chips, data centers, energy, water use) and concentration of power to few major companies as the top two emerging issues in AI governance (R1Q3). In addition, the need for international governance was highlighted given the increasing geopolitical competition over AI. On the other hand, perhaps reflective of the composition of our expert panel, regulation was not seen as a threat to AI innovation but ranked last of the key issues (see also R1Q12). Some respondents also highlighted in qualitative answers that different AI risks are intertwined with each other, given the intersection of AI and other emerging technologies. Moreover, dependence on generative AI was seen by some as potentially leading to erosion of individual agency and decline in cognitive abilities.

---

[2] The same contact list was used across both rounds. While respondents were unaware of each other, the survey was not completely anonymous, as we as researchers needed to validate who had responded twice.
[3] Please see the Appendix A for specific formulation and results of questions R1Q1, R1Q2 and R1Q3.

Below, we have divided the questions into three main clusters, based on their format and type: 1) Landscape of AI policy; 2) Priorities in AI governance; and 3) EU's practical policy options, as a chronological sequence of anticipatory governance, largely following the round structure.

## 4.1. Landscape of AI policy

*Table 3. Results of Likert scale items about probability and desirability (1 = very unlikely/undesirable, and 5 = very likely/desirable) of different AI policy landscape statements. Consensus is based on the highest concentration of numeric scores as per Table 2. See the Appendixes for median, mean and IQR results.*

| Statement | Responses | Consensus |
|---|---|---|
| **R1Q5**: The pace of AI development in the next ten years will be too fast for governments to keep up, resulting in regulatory gaps. | Probability: 3: 18%, 4: 45%, 5: 36% | High: likely to very likely (81%) |
| | Desirability: 1: 48%, 2: 38%, 3: 14% | High: very undesirable to undesirable (86%) |
| **R1Q6**: In the next few years, the development and deployment of increasingly larger and capable AI models, including open-source models will be severely restricted due to safety concerns. | Probability: 1: 5%, 2: 41%, 3: 41%, 4: 9%, 5: 5% | High: neutral to unlikely (82%) |
| | Desirability: 2: 24%, 3: 29%, 4: 38%, 5: 10% | Low: desirable to neutral (67%) |
| **R1Q7**: By the end of the decade, robust instruments of global AI governance have been established through multilateral international treaties or governance bodies (e.g. through the UN). | Probability: 1: 5%, 2: 32%, 3: 36%, 4: 23%, 5: 5% | Low: neutral to unlikely (68%) |
| | Desirability: 2: 5%, 3: 10%, 4: 29%, 5: 57% | High: very desirable to desirable (86%) |
| **R1Q11**: The AI Act's risk-based approach to regulating AI use cases and products will prove itself to be future-proof and appropriate in light of increasingly capable general purpose AI systems. | Probability: 1: 10%, 2: 14%, 3: 43%, 4: 24%, 5: 10% | Low: neutral to likely (67%) |
| | Desirability: 1: 10%, 3: 25%, 4: 40%, 5: 25% | Low: desirable to neutral / very desirable (65%) |
| **R1Q12**: The EU's approach to AI regulation will hinder innovation, leading companies, investors and skilled workers to establish themselves in other parts of the world. | Probability: 1: 14%, 2: 32%, 3: 36%, 4: 14%, 5: 5% | Low: neutral to unlikely (68%) |
| | Desirability: 1: 43%, 2: 33%, 3: 19%, 4: 5% | Medium: very undesirable to undesirable (76%) |
| **R1Q13**: Digital public infrastructure, open-source systems and data/AI commons will form a core part of the success of the EU's approach to technological sovereignty and AI ecosystems. | Probability<br>1: 14%  2: 9%  3: 41%  4: 27%  5: 9%<br><br>Desirability<br>1: 10%  2: 5%  3: 14%  4: 24%  5: 48% | Low: neutral to likely (68%)<br><br>Medium: very desirable to desirable (72%) |
| **R1Q14**: New participatory mechanisms for AI governance, such as citizen panels, will be established in the EU. | Probability<br>1: 9%  2: 36%  3: 27%  4: 18%  5: 9%<br><br>Desirability<br>2: 5%  3: 18%  4: 36%  5: 41% | Low: unlikely to neutral (63%)<br><br>Medium: very desirable to desirable (77%) |
| **R1Q15**: The European AI office will solidify itself as a powerful agency with resources to enforce European AI legislation within the next five years. | Probability<br>1: 5%  2: 18%  3: 41%  4: 23%  5: 14%<br><br>Desirability<br>2: 18%  3: 9%  4: 36%  5: 36% | Low: neutral to likely (64%)<br><br>Medium: desirable to very desirable (72%) |

In terms of the current landscape of AI policy and related expectations, most respondents anticipated that the rapid pace of AI development will outstrip governmental regulatory capacity in the next decade, even if this is very undesirable (R1Q5). One expert highlighted how the AI Act took multiple years to draft and was still seen as rushed and too early by many. Yet, it was also remarked that if one waits for technology to be widely adopted, there is a risk that the market stabilizes to an unsafe equilibrium where intervening is even more difficult (as arguably seen with large language models like ChatGPT, which are hardly compatible with GDPR). In the qualitative comments this was foreseen resulting in major regulatory gaps in line with the classic Collingridge dilemma, while also eliciting more critical perspectives:

> *"The uptake of AI happens at a speed which is higher than our capacity to understand its impacts. The decision making process is slow in democratic systems, as it needs to make sure all societal groups are consulted. In that sense, governments might need to explore agile mechanisms for decision making (regulatory sandboxes or test-beds, agile policy making...) while keeping societal safeguards."* (Respondent 6)

> *"Regulation necessarily lags behind innovation, but this is an old argument and not universally true. Recent research has questioned the truth of this truism, especially in the context of the GDPR and AI Act both of which are technology neutral. I think the purpose of regulation is often misunderstood – it sets out societal values and expectations that new technologies need to align to, not the other way around."* (Respondent 7)

Both the establishment of robust global AI governance mechanisms and new participatory instruments for AI governance within the EU were considered highly desirable by the majority of experts, yet their probability was viewed as considerably lower (R1Q7 & R1Q14). Many thought citizen participation should be carefully applied to the most opportune aspect of AI governance to ensure its effectiveness and to avoid tokenistic participation. Experts thought it was rather unlikely that development of increasingly larger and capable AI models would be significantly restricted by governments for safety reasons (R1Q6), perhaps reflecting the race dynamics of AI: *"The EU AI Act settled on very basic safeguards for the largest open-* *source models and it's one of the most regulation-open regions in the world…"* (Respondent 12). Others questioned the whole safety narrative: *"I believe that the safety perspective on its own is insufficient. Concentrations of power are the major risk that should be addressed through policies shaping open AI development"* (Respondent 18). There was no strong consensus on whether the AI Act's risk-based approach would prove to be future-proof (R1Q11), whether the European AI Office will become a powerful enforcement agency (R1Q15) or whether digital public infrastructure and open-source systems will form a core part of the EU's AI approach (R1Q13), even if all were deemed desirable. One respondent (22) considered the AI Act to be *"Too complex, definitions too broad and ambiguous"* whereas others saw this as a mark of its future-proofness and flexibility:

> *"Initially the high-risk risk-based logic couldn't handle general-purpose AI models well, but the final proposal deals with those models somewhat separately/horizontally as well. There are also some instruments to change the law, which gives me hope it could be future-proof. In the past, updates have been slow, however."* (Respondent 12)

> *"…while additional provisions on advanced, general-purpose AI systems have been included in the AI Act, there are still some concerns around how this is made concrete and actionable, and ultimately this may depend more on the performance of European institutions than on ex ante features of the regulatory framework as written."* (Respondent 13)

### 4.2. Priorities in AI governance

*Table 4. Results of Likert scale items regarding the importance (1 = not important at all, and 5 = very important) of AI governance priorities.*

| R2Q1: How important do you consider the following factors for ensuring that the EU's AI regulation is future-proof and flexible to respond to new developments in AI technologies and the possible regulatory gaps they create?[4] | | Consensus |
|---|---|---|
| 1. Sufficient resources for the AI Office and national authorities for enforcement (total score: 80) | 2: 6%, 3: 11%, 4: 17%, 5: 67% | High: very important to somewhat (84%) |
| 2. Updating and amending the regulation through implementing and delegated acts (total score: 71) | 2: 11%, 3: 11%, 4: 50%, 5: 28% | Medium: somewhat to very important (78%) |
| 3. Technology neutral definition of AI systems (total score: 65) | 1: 11%, 2: 11%, 3: 17%, 4: 28%, 5: 33% | Low: very important to somewhat (61%) |
| 4. Risk-based approach (total score: 65) | 2: 17%, 3: 22%, 4: 44%, 5: 17% | Low: somewhat important to neutral (66%) |
| 5. Standardization process with industry (total score: 64) | 1: 6%, 2: 11%, 3: 28%, 4: 33%, 5: 22% | Low: somewhat important to neutral (61%) |

---

[4] N= for R2Q1 was 18, but 19 for the other table items. This was due to two partial responses, one of which was completely discounted from the results. If these were considered the total score for 'Sufficient resources for the AI Office…' would have been 89 and 81 for 'Updating and amending the regulation…', further validating the findings.

| | | |
|---|---|---|
| 6. National regulatory sandboxes and test-beds (total score: 61) | 2: 22%  3: 28%  4: 39%  5: 11% | Low: somewhat important to neutral (67%) |
| **R2Q2:** How important do you consider each of the following EU digital policies for complementing the implementation and enforcement of the AI Act? | | **Consensus** |
| 1. Digital Services Act (DSA) (total score: 75) | Importance<br>2: 11%  3: 16%  4: 42%  5: 32% | Medium: somewhat to very important (74%) |
| 2. General Data Protection Regulation (GDPR) (total score: 70) | 2: 16%  3: 21%  4: 42%  5: 21% | Low: somewhat to very important / neutral (63%) |
| 2. General Data Protection Regulation (GDPR) (total score: 70) | 2: 16%  3: 21%  4: 42%  5: 21% | Low: somewhat to very important / neutral (63%) |
| 4. Data Act (DA) (total score: 61) | 2: 26%  3: 37%  4: 26%  5: 11% | Low: neutral to somewhat unimportant / important (63%) |
| 5. Data Governance Act (DGA) (total score: 59) | 1: 5%  2: 21%  3: 37%  4: 32%  5: 5% | Low: neutral to somewhat important (69%) |
| **R2Q3:** Which stages of the AI system lifecycle do you think are the most important for the EU to prioritize in its AI regulation in the future? | | **Consensus** |
| 1. AI infrastructure and computing power (total score: 77) | Importance<br>1: 5%  2: 5%  3: 16%  4: 26%  5: 47% | Medium: very important to somewhat (73%) |
| 2. Model development and training (total score: 76) | 2: 11%  3: 11%  4: 47%  5: 32% | Medium: somewhat to very important (79%) |
| 3. Design and data collection phase (total score: 75) | 2: 11%  3: 16%  4: 42%  5: 32% | Medium: somewhat to very important (74%) |
| 4. AI deployment and use (total score: 73) | 2: 16%  3: 11%  4: 47%  5: 26% | Medium: somewhat to very important (73%) |
| 5. Social adoption and proliferation (total score: 72) | 2: 5%  3: 37%  4: 32%  5: 26% | Low: neutral to somewhat important (69%) |
| **R2Q7:** How important do you think the following forums could be for global cooperation on AI governance? | | **Consensus** |
| 1. Cooperation on standards (e.g. ISO/IEC) (total score: 69) | Importance<br>1: 11%  2: 5%  3: 21%  4: 37%  5: 26% | Low: neutral to very important (63%) |
| 2. International agreements and treaties (e.g. CoE, OECD, GPAI, Hiroshima Process) (total score: 68) | 1: 5%  2: 21%  3: 16%  4: 26%  5: 32% | None (less than 60%) |
| 3. A new scientific body for AI governance (e.g. an "IPCC for AI" or a "CERN for AI") (total score: 65) | 1: 11%  2: 32%  3: 5%  4: 11%  5: 42% | None (less than 60%) |
| 4. Network of AI safety institutes for monitoring / evaluation (e.g. US, UK, EU, Japan) (total score: 64) | 1: 11%  2: 16%  3: 21%  4: 32%  5: 21% | None (less than 60%) |
| 5. Wide multistakeholder cooperation through the UN system (total score: 53) | 1: 11%  2: 32%  3: 32%  4: 21%  5: 5% | Low: somewhat unimportant to neutral (64%) |

The expert responses revealed an agreement on several priorities for AI governance, building on the desirable directions outlined in the earlier questions. Foremost was ensuring transparency, fairness, and trustworthiness of AI systems across sectors, in line with the AI Act's use case approach (R1Q8)[5]. This was followed by evaluating safety of advanced AI models, regulating computing power and antitrust policies. Interestingly, multiple respondents highlighted the importance of citizen participation in AI governance in qualitative answers while it ranked last in the quantitative ranking: *"Citizen participation is important, but even more important is to define governance mechanisms in which those participations could feed into"* (Respondent 10). This suggests that while experts view democratic input as desirable, if forced to rank, they prioritize technical regulatory levers, perhaps due to feasibility concerns. Experts noted that the AI Act does not operate in isolation (R2Q2) as a regulatory instrument – it should be implemented in conjunction with the Digital Services Act (DSA), General Data Protection Regulation (GDPR) and to a lesser degree with the Digital Markets Act (DMA). However, it was also noted that the overlaps and differences in legal redress and enforcement authorities between these regulations could result in problematic competence battles. As for the AI lifecycle stages that the EU should prioritize in regulation, AI infrastructure and computing power ranked as a key lever and bottleneck (R2Q3). However, the design, development and deployment of AI systems were seen as equally important and interlinked, with the downstream social adoption slightly less of a priority, perhaps due to its regulatory difficulty.

As for the future-proofness of European AI regulation, the two key factors were deemed to be sufficient enforcement resources for the AI Office and national authorities as well as amendments to the regulation through delegated acts (R2Q1). Some noted that the AI Office not only requires resources but further mandate in terms of foresight, horizon scanning and monitoring of emerging technologies and use cases of AI. Other factors, such as the AI Act's risk-based approach and technology neutrality were not seen quite as important for future-proofness, while still valuable. Some respondents expressed skepticism about the impact of standard-setting and regulatory sandboxes, especially given the monopolizing effects of the large data and compute requirements of modern AI systems. The main expert emphasis was the importance of implementation and enforcement of the regulation:

> *"Risk-based approach in AI regulation should ensure it is future-proof. However, the EU needs to focus strongly on the implementation and supporting companies to be compliant with new*

---

[5] See Appendix A for R1Q8. It is not visualized here due to its different ranking format.

*regulation. Implementation should also enable evaluation of the impact of the regulation and identification of possible regulatory gaps."* (Respondent 14)

*"Regulation will not solve anything on its own. We need very competent authorities that can both enforce and advise."* (Respondent 20)

Despite the issues with the AI Act, respondents noted it is up to the regulators to exercise flexibility and adaptiveness in implementing it. The pivotal role of enforcement culture and competence is also highlighted by the aforementioned inaction regarding ChatGPT's noncompliance with GDPR and EU copyright law. In terms of the most important forums for global cooperation on AI governance (R2Q7), standard-setting through ISO/IEC and international agreements (e.g. CoE, OECD) scored the highest. On the other hand, multistakeholder cooperation through the United Nations scored considerably lower than other options, highlighting doubts about the efficacy and speed of the UN system in contrast to rapid technological development. Nevertheless, the current geopolitical competition and perception of an 'AI race' was not seen as conducive to global governance of AI by experts. As for the most effective points for embedding citizen participation into AI governance (R2Q10)[6], citizens' assemblies and civil society engagement on national visions for AI deployment were mentioned on a national and local level. At the EU level, the respondents saw the European Parliament and the implementation of the AI Act as opportune contexts for participatory oversight, e.g., through standardization work and consultations of the AI Office. Globally participatory AI governance was seen as most challenging, potentially requiring an entirely new AI-focused intergovernmental organization given deficiencies of the UN system. This could be aided by more transparent, informed, deliberative and accountable decision-making processes, which experts saw as AI's largest benefit for democracy (R2Q11).

### 4.3. EU's practical policy options

*Table 5. Results of Likert scale items from strongly disagree (1) to strongly agree (5) for statements about EU's policy options. Consensus based on the highest concentration of numeric scores.*

| Statement | Responses | Consensus |
|---|---|---|
| **R2Q4**: Going forward, the EU's primary focus should be on interpreting and updating existing legislation from other sectors (e.g. health, transportation) into the context of AI instead of introducing new, AI-specific legislation (e.g. AI Liability Directive). | 1: 5%, 2: 11%, 3: 21%, 4: 47%, 5: 16% | Low: agree to neutral (68%) |
| **R2Q5**: The EU's AI policy should continue to rely on industry self-regulation through standardization, conformity assessments and code of practices for general-purpose AI systems. | 1: 35%, 2: 45%, 3: 10%, 4: 10% | High: disagree to strongly disagree (80%) |
| **R2Q6**: The EU should proactively regulate the infrastructure, computing power and chips needed for development of AI systems, instead of merely focusing on the downstream use cases of AI. | 1: 5%, 2: 16%, 3: 21%, 4: 37%, 5: 21% | None (less than 60%) |

---

[6] R2Q10 and R2Q11 are both qualitative, open text field questions not visible here.

| | | Agreement | |
|---|---|---|---|
| **R2Q8**: The effects of EU's AI regulation will be replicated widely internationally due to the 'Brussels effect' – companies will maintain more stringent standards due to economic, legal and technical practicalities of universal standard across jurisdictions. | | 2: 26%  3: 32%  4: 42% | Medium: agree to neutral (74%) |
| **R2Q9**: The primary focus of international AI cooperation should be the risk and safety of advanced 'frontier' AI systems (e.g. Bletchley declaration, safety summits, AI safety institutes). | | 1: 21%  2: 32%  3: 11%  4: 32%  5: 5% | None (less than 60%) |
| **R2Q12**: In order to address the concentration of power in the AI industry, EU's industrial policy for AI should: | | | |
| 1. | Fund and build more digital public infrastructure for AI (e.g. computing power, data commons) | 3: 5%  4: 47%  5: 47% | High: strongly agree to agree (94%) |
| 2. | Enforce more stringent antitrust rules and Digital Market Act policies against gatekeepers to prevent abuses of dominant market position | 1: 5%  3: 5%  4: 37%  5: 53% | High: strongly agree to agree (90%) |
| 3. | Grow and support an EU-wide ecosystem of AI startups and SMEs | 1: 5%  3: 16%  4: 53%  5: 26% | Medium: agree to strongly agree (79%) |
| 4. | Significantly increase joint EU-level funding (e.g. European Sovereignty Fund, STEP) for AI technologies to prevent the distortion of EU internal markets through national state aid | 1: 5%  2: 11%  3: 26%  4: 26%  5: 32% | None (less than 60%) |
| 5. | Increase investments into key sectors of AI value chain (e.g. semiconductors) | 2: 11%  3: 37%  4: 37%  5: 16% | Medium: neutral to agree (74%) |
| 6. | Direct significantly more investments into open-source AI systems | 1: 5%  2: 16%  3: 16%  4: 47%  5: 16% | Low: agree to neutral / agree to strongly agree (63%) |
| 7. | Focus on growing and supporting its own large scale AI companies | 1: 16%  2: 11%  3: 53%  4: 21% | Medium: neutral to agree (74%) |

In terms of practical AI regulation, there was an inclination towards updating existing sector-specific legislation to the context of AI, although the AI Liability Directive or similar regulation was still seen as an important AI-specific addition by respondents (R2Q4). Consistent with the other findings, experts strongly disagreed that AI policy should primarily rely on industry self-regulation (R2Q5): *"Competition and market dynamics make self-regulation wholly unsuitable. We've seen this approach fail in the UK - where model providers don't provide early access for testing, despite committing to do so"* (Respondent 25). There was relative agreement on the need for the EU to proactively regulate the underlying AI infrastructure, including computing power and chips, even if this might be practically challenging (R2Q6). However, opinions were divided on whether the 'Brussels effect' would lead to a wider adoption of EU's AI regulation globally (R2Q8) and whether international AI cooperation should primarily focus on the risks and safety of advanced 'frontier' AI systems (R2Q9). As noted by one respondent (14): *"Safety discourse is partly reductive in the context of AI but also it is the framework that can enable international cooperation."*

As for the most important gaps to be addressed in the EU's approach to AI regulation (R1Q16)[7], the concentration of power to the largest US-based AI companies was seen as the most pressing (55%), followed by the lack of citizen participation instruments, insufficient intergovernmental bodies for AI governance, and inadequate investments into AI commons (all 45%). In addressing the concentration of power within the AI industry (R2Q12), experts strongly supported investing in digital public infrastructure for AI and enforcing more stringent antitrust rules against market gatekeepers. Multiple respondents called for an outcome-based industrial strategy for AI based on clear endpoints. Some experts felt that given the AI industry's tendency towards monopolies, expecting the market to self-correct with minimal interventions (like providing compute for start-ups) is insufficient. Instead, they suggested setting high regulatory standards, akin to the FDA in the US, or breaking up monopolies and replacing them with a public-interest model (e.g., CERN for AI).

> *"We need a proper CERN for AI, similar to ESA, with significant funding (100 billion EUR for 10 years) and significant flexibility. With the small and fragmented investments the EU is making today, they will fail. The member states are equally bad. Too little, too late. Unfortunately."* (Respondent 20)

> *"The key, I think, is to extract Europe from the AI race mentality. Rather than running after the others, we should ask ourselves what European societies need to progress, for their citizens. Sometimes that may involve AI, but in many aspects, it may not. So less of a blind embrace of tech solutions for problems that have nothing to do with a lack of technology (polarisation, distrust, inequality, environmental degradation, etc.)."* (Respondent 9)

There was also support for directing more investments into open-source AI systems and other joint EU-level AI funding, but respondents noted this does not necessarily limit market power. To address the power concentration at the infrastructure layer of AI, it might be better to invest in open-source scaffolding infrastructures. While supporting European AI startups and SMEs was well received, their reliance on the infrastructure of hyperscalers was still seen as a challenge, which could lead to public investments being commercialized by foreign technology companies. Growing European large scale AI companies was evaluated less favourably:

> *"I don't really see why a European competitor necessarily offers better outcomes for citizens. They'd be subject to the same competitive dynamics as US companies. A CERN for AI seems more obviously in the public interest to me."* (Respondent 25)

Investments into the key parts of the AI value chain were seen as generally desirable to ensure digital sovereignty, but some questioned the focus on semiconductors and instead saw cloud computing as a more realistic avenue. Others pointed out pouring tens of billions of euros into chip foundries in the EU would require a broader industrial policy strategy that leverages European assets such as ASML more strategically.

---

[7] Please see the Appendix A for details of R1Q16.

Nevertheless, these industrial policy efforts were still widely seen to require regulation and stringent antitrust rules against gatekeeper companies to prevent abuses of dominant market position. Overall, the expert responses indicate that the EU has the potential to exercise meaningful agency over the future of AI governance, but this requires vigorous enforcement of its existing competition, consumer protection, and copyright laws, including GDPR.

5. Discussion

Several substantive insights for AI policy emerge from the Policy Delphi, particularly from the perspective of anticipatory governance and the EU's ambition to ensure its AI regulation remains future-proof. First, there exists a significant desirability-probability gap in regulation: the most desirable AI policy options are perceived as some of the least feasible and unrealistic (see section 4.1). Most notably, there is a universal consensus on the need for robust international coordination and citizen participation in AI governance, which contrasts sharply with their perceived low probability. Similar pessimism about implementing desirable policy interventions has been observed by Stahl et al. (2023), although in their Delphi study formal regulation and citizens' juries were one of the least supported mechanisms of AI governance. Our findings suggest a tension between the desire for proactive regulatory oversight and the practical difficulty for regulation to keep pace with rapid AI development. For instance, the Policy Delphi respondents expressed strong support for digital public infrastructure and antitrust measures to address power concentration in AI, yet simultaneously predicted governments would struggle to keep pace with AI development. Moreover, the experts stressed the EU's own agency to choose its direction in the 'AI race' and reject industry self-regulation, rather than necessarily creating its own domestic AI giants (see Mügge, 2024). Yet, given the desirability-probability gap, there are reasons to doubt the feasibility of this desired proactive industrial policy and democratic oversight in AI governance. Indeed, the experts seem to expect that the speed of AI deployment will favour technocratic, market-led governance structures, following the pacing problem (Downes, 2009).

The tension between desirability and feasibility of policies is of course not unique to AI governance. European Commission's former president Jean-Claude Juncker has been reported as saying *"We all know what to do, we just don't know how to get re-elected once we've done it."* (Buti et al., 2009), referring to the trade-off between financially responsible long-term reforms and short-term electoral incentives. Similarly, beneficial long-term decisions like technology regulation may be politically unpopular due to their short-term costs (see Boston, 2017). However, AI's rapid progress and uncertainty challenges the normal, electoral cycle-based timeframes of politics. AI governance is characterized by concentration of power to private technology giants, rather than states, which diminishes the temporal effect of elections and parliamentary calendars. As such, operational logic of powerful technology companies, their CEOs and technocratic experts heavily influences timescales in AI governance by pushing for certain temporal narratives of the technology. This is particularly salient in the discourse on race towards AGI or superintelligence, which many industry leaders say is mere years away (Ó hÉigeartaigh, 2025). These expectations are also reflected in the debate around safety of advanced AI systems, which many experts in our survey approached with skepticism. Despite similar numerical estimates of AI risks, the qualitative answers revealed a division between respondents focused on near-to-mid-term societal risks and those concerned with long-term existential safety risks. Arguably, these narratives are a significant contributor to the desirability-probability gap, as the more desirable governance options could entail 'losing the AI race', thereby limiting the perceived policy options.

Second, the enforcement and implementation of existing regulation emerged as key linchpin for anticipatory AI governance – regulation is only as good as it is implemented. In contrast, one of the most well-known mechanisms of future-proof regulation, technology neutrality, was not seen nearly as relevant by the Policy Delphi respondents: *"Technology neutral regulation is desirable but a technology neutral definition of 'AI' may be a red herring. A risk-based approach brackets out values and concerns that are not readily expressed in the language of 'risks'."* (Respondent 26). Another respondent emphasized courts can already interpret the law flexibly under 'teleological interpretation' to keep up with technological progress. Similarly, regulatory sandboxes as spaces for technological and regulatory experimentation before full legal compliance ranked last out of the future-proof regulatory options, despite the EU increasingly applying them for anticipatory governance (Umbach, 2024, 417). Experts also discounted co-regulatory schemes, like industry-driven standards or codes of practice that are usually taken to aid regulatory flexibility and future-proofness (Ranchordás & van 't Schip, 2020). However, this must be contextualized by the panel's composition, which favored academic, policymaker and civil society perspectives. In contrast, experts showed some willingness to regulate access to AI infrastructure and compute, which could be deemed anticipatory by targeting the upstream lifecycle of AI systems. But such a heavy-handed approach might also quickly become obsolete if the technical paradigm shifts (e.g., as models become more efficient), pointing to a potential research direction for future AI Policy Delphi's. Nevertheless, above all else experts highlighted sufficient resources for authorities to enforce the regulation and update it accordingly in response to technological progress.

This focus on implementation of legislation supports Ibáñez Colomo's (2022) argument that enduring political commitment to the regulation and its enforcement are the main factors for future-proofness. The failure or success of future-proof regulation primarily hinges on the intertemporal consistency of legislatures, rather than on technical design choices like technology neutrality. For instance, while the EU's telecommunications regulation has managed to adapt to transformations of the technological landscape so far, the EU legislature's commitment to flexible and adaptable implementation of it has waned over time (Ibáñez Colomo, 2022). In other words, the regulation's robustness over time is less reliant on technical drafting of statutes or masterfully crafting its scope, but more so on the commitment of policymakers to upholding the regulation as originally understood. This dynamic was clearly visible in the Policy Delphi, where multiple respondents criticized the European Commission's weak enforcement and minimal amendments to the GDPR (see Arnal, 2025). The finding suggests the agile capacity of the institutions, like resources for enforcement, require more emphasis in anticipatory governance to complement foresight. It is specifically this regulatory commitment that is now being tested with the EU's shifting policy focus towards economic competitiveness, characterized by deregulation and simplification of existing legislation such as the AI Act (European Commission, 2025).

Third, what are we to make of the value of the Policy Delphi method in the governance of rapidly evolving technologies like AI? Overall, the results showcase the contribution of the Policy Delphi to some of the central puzzles of future-proof and anticipatory governance, such as the role of technology neutrality. As one of the few Delphi studies focused specifically on AI governance (see Alon et al. 2025), our study sheds light on the tension between probable and desirable AI policy pathways and the proactive governance measures relating to them. The probability-desirability distinction can expose underlying roadblocks to anticipatory governance of AI, such as the required industrial policy investments. Moreover, the Policy Delphi can help assess whether there is sufficient political will and institutional capacity for binding legislative frameworks, or whether soft-law mechanisms would prove more future-proof. Our results showcase a preference for enforceable regulation to tackle the emerging risks of AI and power concentration, even if softer, voluntary measures were viewed as more probable given political realities. The Policy Delphi survey was conducted in a turbulent time period where the AI Act's practical implementation was still being finalized. Yet, several findings of the survey have since been corroborated, such as the lack of concrete restrictions on large AI models due to safety reasons. Overall, the Policy Delphi exhibited greater consensus or at least neutrality regarding policy claims than anticipated. For the more controversial questions, like the divergent definitions of AI risks and safety, the qualitative responses proved valuable in uncovering nuances hidden by numerical Likert scores.

Our study comes with certain limitations. Its geographical scope is confined to AI governance within the EU, potentially limiting the generalizability of the results to national and global context. As noted by Stahl et al. (2023), the expert-centric nature of the Policy Delphi can also limit the applicability of findings to the general public. Furthermore, the final panel consisted of 29 experts with fewer respondents from the AI industry compared to the other categories. This skew in panel composition might have contributed to a greater faith in formal regulation and industrial policy and should be considered when interpreting some of the results. However, the panel composition does not appear as relevant for the findings on the importance of implementation for future-proof AI regulation, as the private sector also shares an equal interest in legal certainty. As for the survey items, the ranking questions in round one proved difficult to answer for the experts, limiting their utility. Otherwise, the risks of inconsistently applying the Policy Delphi method were mitigated through adherence to established guidelines and thorough methodological documentation (see De Loë et al., 2016). The reliance on numerical Likert scale scores was purposefully mitigated by including an open-ended response option with each question. Given these measures, we maintain that the study provides valuable insights for more anticipatory AI governance.

## 6. Conclusion

This article has explored the value of the Delphi method for anticipatory AI governance and future-proof regulation. To our knowledge, this is the first Policy Delphi to specifically assess the governance of AI itself, rather than forecasting AI development in specific areas. We have argued that Policy Delphi's focus on policy disagreement makes it uniquely suited to the rapidly evolving and uncertain landscape of AI governance. Based on our results, this approach can valuably inform more anticipatory governance of emerging technologies, such as AI. Our analysis revealed two substantive, interrelated insights that challenge current EU policy. First, there is an important desirability-probability gap in AI governance: while experts strongly supported citizen participation and international governance of AI, they simultaneously viewed these desirable interventions as most improbable in the face of rapid technological development. This suggests that the pacing problem (Downes, 2009) poses a real barrier for democratic oversight of AI. Second, contrary to the EU's emphasis on technology neutral law and regulatory sandboxes, expert responses suggest that future-proof AI regulation depends primarily on the institutional capacity of bodies like the AI Office to enforce and update the rules. As such, future-proofness appears to have less to do with the technical details or scope of the law, and more with political commitment and resources for its practical implementation (see Ibáñez Colomo, 2022). Regulation is only as good as it is implemented. This warning is particularly salient in the current European political landscape, where the Digital Omnibus (European Commission, 2025) potentially strips authorities of some of the very enforcement tools required to make the AI Act future-proof.

**Acknowledgements**

This research was conducted as part of the KT4D project, which has received funding from the European Union's Horizon Europe program under grant agreement No 101094302. Earlier versions of this paper have been presented and received valuable feedback at the Futures Conference 2025 in Turku, Finland and the 2025 IPSA World Congress of Political Science in Seoul, South Korea. We also want to thank Tuukka Puonti for their help with the data analytics and visualization.

# Appendixes

## Appendix A. Policy Delphi round 1 questionnaire.

- Open for three weeks from 16 April to 7 May 2024.
- N= 21-22, while being 20 for one question (R1Q11).
- All the questions also included an optional *"Your reasoning and other relevant comments"* part, not visible here. Similarly, answers to the open questions are not included here.
- Results are shown in terms of median, mean (standard deviation) and interquartile range (IQR), aside from questions R1Q2 and R1Q16, which use a different metric.

| Question | Responses | Median | Mean (SD) | IQR |
|---|---|---|---|---|
| **R1Q1** Within the next two decades how large of a societal impact do you expect development of AI systems to have? *Assess the transformativeness and risks associated with this development on a scale of 1-5.* | Transformativeness<br><br>*1 = Negligible, 2 = Minor, 3 = Moderate, 4 = Significant, 5 = Transformative* | 4 | 4.23 (0.67) | 1.0 |
| | Risks involved<br><br>*1 = Negligible, 2 = Minor, 3 = Moderate, 4 = Significant, 5 = Extreme* | 4 | 3.95 (0.65) | 0.0 |
| **R1Q2** Which kind of risks of AI do you see most prominent for democratic societies? *Select three at most.* | Erosion of knowledge and epistemic agency of citizens | Count: 7 | Pct: 31.8% | - |
| | Concentration of power to AI companies | 12 | 54.5% | - |
| | Disinformation, filter bubbles and electoral microtargeting | 10 | 45.5% | - |
| | Fairness, biases and discrimination | 9 | 40.9% | - |
| | Opaque and non-transparent AI systems | 4 | 18.2% | - |
| | Privacy and surveillance | 5 | 22.7% | - |
| | Job displacement, unemployment and inequality | 7 | 31.8% | - |
| | Safety and security of advanced AI systems | 5 | 22.7% | - |
| **R1Q3** Rank the following issues from 1–7 (with 1 being the most important) in terms of how likely they are to become this decade's key issues or trends in AI governance. | Ethical and trustworthy AI systems in general | 5 | 4.33 (1.96) | 3.0 |
| | Ensuring that regulation does not stifle AI innovation | 6 | 5.10 (2.15) | 3.0 |
| | Intensified geopolitical competition | 5 | 4.33 (2.03) | 3.0 |
| | Material AI infrastructure (chips, data centers, energy, water use) | 3 | 3.38 (1.80) | 2.0 |
| | Concentration of power and infrastructure to few major companies | 3 | 3.52 (1.60) | 2.0 |
| | Misalignment and safety risks of advanced AI models | 4 | 3.76 (1.76) | 2.0 |
| | The need for robust global and international AI governance | 3 | 3.57 (2.29) | 4.0 |
| **R1Q4** OPEN QUESTION: Are there other AI development trends that you think require consideration? | | | | |
| **R1Q5**: The pace of AI development in the next ten years will be too fast for governments to keep up, resulting in regulatory gaps.  *Assess how probable and how desirable this statement is on a scale of 1–5.* | Probability  *1 = very unlikely, 2 = unlikely, 3 = neutral, 4 = likely, 5 = very likely* | 4 | 4.18 (0.73) | 1.0 |
| | Desirability  *1 = very undesirable, 2 = undesirable, 3 = neutral, 4 = desirable, 5 = very desirable* | 2 | 1.67 (0.73) | 1.0 |
| **R1Q6**: In the next few years, the development and deployment of increasingly larger and capable AI models, including open-source models will be severely restricted due to safety concerns. | Probability | 3 | 2.68 (0.89) | 1.0 |
| | Desirability | 3 | 3.33 (0.97) | 1.0 |
| **R1Q7**: By the end of the decade, robust instruments of global AI governance have been established through multilateral international treaties or governance bodies (e.g. through the UN). | Probability | 3 | 2.91 (0.97) | 1.75 |
| | Desirability | 5 | 4.38 (0.86) | 1.0 |
| **R1Q8:** Evaluate the importance of the following aspects of AI governance by ranking them from 1 to 5, with 1 being the most important. Consider the implications each aspect has on the development, deployment, and societal impact of AI systems.[8] | Ensuring transparency, fairness, and trustworthiness in AI systems in different sectoral use cases (total score: 59) | 2 | 2.81 (1.75) | 4.0 |
| | Regulating access to computing power and the use of data (total score: 63) | 3 | 3.0 (1.34) | 2.0 |
| | Addressing the concentration of power in the AI landscape through antitrust and open-source policies (total score: 64) | 3 | 3.05 (1.47) | 2.0 |
| | Guaranteeing the safety of advanced AI models through evaluations and restrictions on model development (total score: 63) | 3 | 3.0 (1.10) | 0.0 |
| | Ensuring citizen participation in shaping AI governance policies (total score: 66) | 3 | 3.14 (1.49) | 3.0 |
| **R1Q9** OPEN QUESTION: Are there other emerging issues, gaps or major risks related to the governance of AI that you think are important to consider? | | | | |
| **R1Q10** How important is each of the following for the future of AI governance in the EU (1–5)?[9]  *1 = Not important  2 = Slightly important* | Technological sovereignty (total score: 65) | 3 | 3.10 (1.51) | 2.0 |
| | European values and democratic governance of AI (total score: 76) | 4 | 3.62 (1.56) | 3.0 |
| | Global competitiveness and innovations (total score: 74) | 4 | 3.52 (1.33) | 2.0 |

---

[8] Note R1Q8 (like R1Q3) had 1 as the most important rank. Lower score therefore signals more importance. If the scale was inverted 'Ensuring transparency, fairness, and trustworthiness' would score 67, and 'Ensuring citizen participation in shaping AI governance policies' 60 in total.

[9] R1Q10 was excluded from the results section as the respondents interpreted it differently, making the data unreliable. Some approached it as a ranking question, only assigned each number once.

| | | | | |
|---|---|---|---|---|
| 3 = Moderately important<br>4 = Important<br>5 = Very important | Ethics, transparency and trustworthiness (total score: 71) | 4 | 3.38 (1.32) | 1.0 |
| | AI safety and security concerns (total score: 73) | 4 | 3.48 (1.25) | 1.0 |
| **R1Q11**: The AI Act's risk-based approach to regulating AI use cases and products will prove itself to be future-proof and appropriate in light of increasingly capable general purpose AI systems.<br><br>*Assess how probable and how desirable this statement is on a scale of 1–5.* | Probability<br><br>*1 = very unlikely, 2 = unlikely, 3 = neutral, 4 = likely, 5 = very likely* | 3 | 3.10 (1.09) | 1.0 |
| | Desirability<br><br>*1 = very undesirable, 2 = undesirable, 3 = neutral, 4 = desirable, 5 = very desirable* | 4 | 3.70 (1.17) | 1.25 |
| **R1Q12**: The EU's approach to AI regulation will hinder innovation, leading companies, investors and skilled workers to establish themselves in other parts of the world. | Probability | 3 | 2.64 (1.05) | 1.0 |
| | Desirability | 2 | 1.86 (0.91) | 1.0 |
| **R1Q13**: Digital public infrastructure, open-source systems and data/AI commons will form a core part of the success of the EU's approach to technological sovereignty and AI ecosystems. | Probability | 3 | 3.09 (1.15) | 1.0 |
| | Desirability | 4 | 3.95 (1.32) | 2.0 |
| **R1Q14**: New participatory mechanisms for AI governance, such as citizen panels, will be established in the EU. | Probability | 3 | 2.82 (1.14) | 1.75 |
| | Desirability | 4 | 4.14 (0.89) | 1.0 |
| **R1Q15**: The European AI office will solidify itself as a powerful agency with resources to enforce European AI legislation within the next five years. | Probability | 3 | 3.23 (1.07) | 1.0 |
| | Desirability | 4 | 3.91 (1.11) | 1.75 |
| **R1Q16**: In your view, what are the most important gaps in the EU's approach to AI regulation that should be addressed during this decade?<br><br>*Please choose and rank 3, with 1 being the most important.* | a) Too stringent AI regulation hindering innovation, global competitiveness and technological sovereignty of the EU | Count: 3 | Weighted [10] score: 7 | Pct: 14% |
| | b) Pace of technological development too fast for regulation to keep up | 5 | 11 | 23% |
| | c) The risk-based approach adopted in the EU AI Act falls short of protecting people's fundamental rights | 7 | 14 | 32% |
| | d) Concentration of power to the largest US-based AI companies and threat of regulatory capture by industry | 12 | 23 | 55% |

---

[10] Weighted score is calculated as 1 = 3 points, 2 = 2 points, and 3 = 1 point.

|  | | | | |
|---|---|---|---|---|
| | e) The lack of instruments for civil society and citizen participation in AI governance harms democratic decision-making and trustworthiness of the technology | 10 | 21 | 45% |
| | f) Without a global or intergovernmental bodies for AI governance, national or regional efforts will ultimately be insufficient | 10 | 20 | 45% |
| | g) Lack of investments into digital public infrastructure and other AI commons | 10 | 19 | 45% |
| | h) Inadequate consideration of the environmental footprint of AI systems (e.g. carbon and water footprint) | 2 | 2 | 9% |
| | i) Insufficient emphasis on safety and security of advanced AI models | 5 | 12 | 23% |
| | j) Unclear liability for AI harms between providers and users of general-purpose AI systems | 2 | 3 | 9% |

**R1Q17** OPEN QUESTION In terms of policy, governance and regulation, what steps should be taken to address these gaps? *Please answer at least to the three gaps you chose in the question above.*

**R1Q18** OPEN QUESTION**:** Are there other pressing questions, dynamics or regulatory priorities related to European AI policy and governance that you think are important to consider, e.g. promising policy approaches from other regions?

**Appendix B. Policy Delphi round 2 questionnaire.**
- Open for three weeks from 28 May to 18 June 2024.
- N=19 aside from R2Q1, where one partial response was removed.
- As in the first round, all questions also included "Your reasoning and other relevant comments" section, but the results of these or open questions are not included here.
- Results are shown in terms of median, mean (SD) and interquartile range (IQR).

| Question | Responses | Median | Mean (SD) | IQR |
|---|---|---|---|---|
| **R2Q1**: How important do you consider the following factors for ensuring that the EU AI regulation is future-proof and flexible to respond to new developments in AI technologies and the possible regulatory gaps they create?<br><br>*1 = not important at all,*<br>*2 = somewhat unimportant,*<br>*3 = neutral,*<br>*4 = somewhat important,*<br>*5 = very important* | Technology neutral definition of AI systems (total score: 65) | 4 | 3.61 (1.38) | 2.0 |
| | Risk-based approach (total score: 65) | 4 | 3.61 (0.98) | 1.0 |
| | Standardization process with industry (total score: 64) | 4 | 3.56 (1.15) | 1.0 |
| | Updating and amending the regulation through implementing and delegated acts (total score: 71) | 4 | 3.94 (0.94) | 0.75 |
| | National regulatory sandboxes and test-beds (total score 61) | 3.5 | 3.39 (0.98) | 1.0 |
| | Sufficient resources for the AI Office and national authorities for enforcement (total score: 80) | 5 | 4.44 (0.90) | 1.0 |
| **R2Q2:** How important do you consider each of the following EU digital policies for complementing the implementation and enforcement of the AI Act? | General Data Protection Regulation (GDPR) (total score: 70) | 4 | 3.68 (1.00) | 1.0 |
| | Digital Services Act (DSA) (total score: 75) | 4 | 3.95 (0.97) | 1.5 |
| | Digital Markets Act (DMA) (total score: 69) | 4 | 3.63 (0.90) | 1.0 |
| | Data Act (DA) (total score: 61) | 3 | 3.21 (0.98) | 1.5 |
| | Data Governance Act (DGA) (total score: 59) | 3 | 3.11 (0.99) | 1.5 |
| **R2Q3:** Which stages of the AI system lifecycle do you think are the most important for the EU to prioritize in its AI regulation in the future? | AI infrastructure and computing power (total score: 77) | 4 | 4.05 (1.18) | 1.5 |
| | Design and data collection phase (total score: 75) | 4 | 3.95 (0.97) | 1.5 |
| | Model development and training (total score: 76) | 4 | 4.0 (0.94) | 1.0 |
| | AI deployment and use (total score: 73) | 4 | 3.84 (1.01) | 1.0 |
| | Social adoption and proliferation (total score: 72) | 4 | 3.79 (0.92) | 1.5 |
| **R2Q4:** Going forward, the EU's primary focus should be on interpreting and updating existing legislation from other sectors (e.g. health, transportation) into the context of AI instead of introducing new, AI-specific legislation (e.g. AI Liability Directive). | Agreement<br><br>*1= Strongly disagree,*<br>*2 = Disagree,*<br>*3 = Neither agree nor disagree,*<br>*4 = Agree,*<br>*5 = Strongly agree* | 4 | 3.58 (1.07) | 1.0 |
| **R2Q5:** The EU's AI policy should continue to rely on industry self-regulation through standardization, | Agreement | 2 | 1.95 (0.97) | 1.0 |
| conformity assessments and code of practices for general-purpose AI systems. | | | | |
| **R2Q6:** The EU should proactively regulate the infrastructure, computing power and chips needed for development of AI systems, instead of merely focusing on the downstream use cases of AI. | Agreement | 4 | 3.53 (1.17) | 1.0 |
| **R2Q7:** How important do you think the following forums could be for global cooperation on AI governance?<br><br>*1 = not important at all,*<br>*2 = somewhat unimportant,*<br>*3 = neutral,*<br>*4 = somewhat important,*<br>*5 = very important* | A new scientific body for AI governance (e.g. an "Intergovernmental Panel for Climate Change for AI" or a "CERN for AI") | 4 | 3.42 (1.57) | 3.0 |
| | Cooperation on standards (e.g. ISO/IEC) | 4 | 3.63 (1.26) | 1.5 |
| | Network of AI safety institutes for monitoring/evaluation (following the lines of those established in e.g. US, UK, EU, Japan) | 4 | 3.37 (1.30) | 1.5 |
| | International agreements and treaties (e.g. CoE, OECD, GPAI, the Hiroshima Process) | 4 | 3.58 (1.30) | 2.5 |
| | Wide multistakeholder cooperation through the UN system | 3 | 2.79 (1.08) | 1.5 |
| **R2Q8:** The effects of EU's AI regulation will be replicated widely internationally due to the 'Brussels effect' – companies will maintain more stringent standards due to economic, legal and technical practicalities of universal standard across jurisdictions. | Agreement<br><br>*1= Strongly disagree,*<br>*2 = Disagree,*<br>*3 = Neither agree nor disagree,*<br>*4 = Agree,*<br>*5 = Strongly agree* | 3 | 3.16 (0.83) | 1.5 |
| **R2Q9:** The primary focus of international AI cooperation should be the risk and safety of advanced 'frontier' AI systems (e.g. Bletchley declaration, safety summits, AI safety institutes). | Agreement | 2 | 2.68 (1.29) | 2.0 |
| **R2Q10:** OPEN QUESTION: What do you think are the most effective points for embedding democratic input and citizen participation into AI governance on the national, EU, and global levels? (e.g. standardization bodies, defining high-risk systems, alignment assemblies) | | | | |
| **R2Q11:** OPEN QUESTION: What do you see as the most positive use cases of AI, especially in promoting democratic participation (e.g. transparent decision-making, democratization of education, AI for sustainability)? | | | | |
| **R2Q12:** In order to address the concentration of power in the AI industry, EU's industrial policy for AI should: | 12.1 Direct significantly more investments into open-source AI systems | 4 | 3.53 (1.12) | 1.0 |
| | 12.2 Fund and build more digital public infrastructure for AI (e.g. computing power, data commons) | 4 | 4.42 (0.61) | 1.0 |
| *1= Strongly disagree,*<br>*2 = Disagree,*<br>*3 = Neither agree nor disagree,*<br>*4 = Agree,*<br>*5 = Strongly agree*<br><br>*Why do you agree or disagree? -for each option.* | 12.3 Focus on growing and supporting its own large scale AI companies | 3 | 2.79 (0.98) | 0.5 |
| | 12.4 Grow and support an EU-wide ecosystem of AI startups and SMEs | 4 | 3.95 (0.97) | 0.5 |
| | 12.5 Increase investments into key sectors of AI value chain (e.g. semiconductors) | 4 | 3.58 (0.90) | 1.0 |
| | 12.6 Significantly increase joint EU-level funding (e.g. European Sovereignty Fund, STEP) for AI technologies to prevent the distortion of EU internal markets through national state aid | 4 | 3.68 (1.20) | 2.0 |
| | 12.7 Enforce more stringent antitrust rules and Digital Market Act policies against gatekeepers to prevent abuses of dominant market position | 5 | 4.32 (1.00) | 1.0 |
| **OPEN QUESTION**: Thank you for your answers. Are there other major AI policy and governance responses that you would still like to highlight, that haven't been covered enough in this survey? | | | | |

**Appendix C. Anonymized list of expert respondents**

| Respondent number | Expert category | Round participation |
|---|---|---|
| Respondent 1 | NGOs and think tanks | 1 |
| Respondent 2 | NGOs and think tanks | 1 and 2 |
| Respondent 3 | Policymakers (EC and EP) | 1 |
| Respondent 4 | Policymakers (EC and EP) | 1 and 2 |
| Respondent 5 | Academics (AI governance, law and ethics) | 1 and 2 |
| Respondent 6 | Policymakers (EC and EP) | 1 and 2 |
| Respondent 7 | Academics (AI governance, law and ethics) | 1 and 2 |
| Respondent 8 | Academics (AI governance, law and ethics) | 1 |
| Respondent 9 | Academics (AI governance, law and ethics) | 1 and 2 |
| Respondent 10 | Academics (AI governance, law and ethics) | 1 |
| Respondent 11 | Policymakers (EC and EP) | 1 |
| Respondent 12 | NGOs and think tanks | 1 and 2 |
| Respondent 13 | Academics (AI governance, law and ethics) | 1 and 2 |
| Respondent 14 | Policymakers (EC and EP) | 1 and 2 |
| Respondent 15 | NGOs and think tanks | 1 |
| Respondent 16 | Academics (AI governance, law and ethics) | 1 |
| Respondent 17 | Industry | 1 |
| Respondent 18 | NGOs and think tanks | 1 and 2 |
| Respondent 19 | NGOs and think tanks | 1 |
| Respondent 20 | Academics (AI governance, law and ethics) | 1 and 2 |
| Respondent 21 | NGOs and think tanks | 1 and 2 |
| Respondent 22 | Industry | 1 |
| Respondent 23 | Industry | 2 |
| Respondent 24 | NGOs and think tanks | 2 |
| Respondent 25 | NGOs and think tanks | 2 |
| Respondent 26 | Academics (AI governance, law and ethics) | 2 |
| Respondent 27 | NGOs and think tanks | 2 |
| Respondent 28 | Policymakers (EC and EP) | 2 |
| Respondent 29 | Policymakers (EC and EP) | 2 |